\documentclass[]{interact}
\usepackage{hyperref}
\hypersetup{
  pdfproducer={},
  pdfcreator={},
}
\usepackage{amsmath}
\usepackage{graphicx}
\usepackage{subcaption}
\usepackage{breqn}
\newcommand{\ulq}{\mathopen{\textnormal{\textquotesingle}}}
\newcommand{\urq}{\mathclose{\textnormal{\textquotesingle}}}
\usepackage{multirow}
\usepackage[numbers]{natbib}

\begin{document}
\title{Evaluation Of P300 Speller Performance Using Large Language Models Along With Cross-Subject Training}

\author{
\name{
Nithin Parthasarathy\textsuperscript{a}, James Soetedjo\textsuperscript{b}, Saarang Panchavati\textsuperscript{c}, Nitya Parthasarathy\textsuperscript{d}, Corey Arnold\textsuperscript{c}, Nader Pouratian\textsuperscript{e} and William Speier\textsuperscript{c}
}
\affil{\textsuperscript{a}Dept of Comp. Science,University of Illinois at Urbana Champaign;
\textsuperscript{b}Department of Bioengineering, University of Washington;
\textsuperscript{c}Department of Radiological Sciences, UCLA;
\textsuperscript{d}PE Investments, Boston, MA, (formerly at Department of Computer Science, MIT);
\textsuperscript{e}Department of Neurological Surgery, University of Texas, Southwestern, Dallas
}
}
\maketitle

\begin{abstract}
Amyotrophic lateral sclerosis (ALS), a progressive neuromuscular degenerative disease, severely restricts patient communication capacity within a few years of onset, resulting in a significant deterioration of quality of life. The P300 speller brain-computer interface (BCI) offers an alternative communication medium by leveraging a subject's EEG response to characters traditionally highlighted on a character grid on a graphical user interface (GUI). A recurring theme in P300-based research is enhancing performance to enable faster subject interaction. This study builds on that theme by addressing key 
limitations, particularly in the training of multi-subject 
classifiers, and by integrating advanced language models to 
optimize stimuli presentation and word prediction, thereby 
improving communication efficiency. A common challenge in training efficient P300-based multi-subject classifiers is 
tackled through the introduction of novel {\it across} subject classifiers. Furthermore, various advanced large 
language models such as Generative Pre-Trained Transformer (GPT-2), BERT, and BART, alongside Dijkstra's algorithm, are utilized
to optimize stimuli and provide word completion choices based on the spelling history. In addition, a multi-layered smoothing approach is applied to allow for out-of-vocabulary 
(OOV) words. By conducting extensive simulations based on randomly sampled EEG data from subjects, we show substantial speed improvements in typing passages that include rare and out-of-vocabulary (OOV) words, with the extent of improvement varying depending on the language model utilized. The gains through such character-level interface optimizations are approximately $ 10\% $, and GPT2 for multi-word prediction provides gains of around $ 40\% $. In particular,
some large language models achieve performance levels within $ 10\% $ of the theoretical performance limits established in this study. In addition, both within and across subjects,
training techniques are explored, and speed improvements are shown to hold in both cases.\footnote{This study was 
conducted with the principles embodied in the Declaration of Helsinki and in accordance with local statutory requirements. 
The study was approved by the UCLA Institutional review board (IRB\#11-002062) and all subjects provided their written informed consent to participate in the study. The code and data are available at \url{https://github.com/nithinparthasarthy/P300Speller.git}} 
\end{abstract}

\section{Introduction}
Amyotrophic lateral sclerosis (ALS) is a progressive neurodegenerative disease affecting motor neurons in the cerebral cortex that severely impairs patients' lives \cite{ALS}. ALS patients currently have a means to communicate through noninvasive brain-computer interfaces (BCI) \cite{gao}-\cite{mcfarland2} which allow direct translation of electrical, magnetic or metabolic brain signals into control commands of external devices. The P300 speller is a common BCI communication interface that presents stimuli to produce an evoked response. In this technique, a character matrix is presented to the subject and their EEG response (more specifically, the P300 evoked response potential or ERP) which varies based on the highlighted display characters is recorded, processed and interpreted.\cite{mcfarland2}  

Farwell and Donchin introduced the classical paradigm for the P300-based BCI speller in 1988 \cite{farwell}. The Row-Column (RC) paradigm is the most popular speller format, which consists of a matrix of $ 6\times 6 $ characters. In this paper, we will refer to this matrix or GUI as a {\it flashboard}. This flashboard is presented on the computer screen, and the row and columns are typically flashed/highlighted in a random order, with the subject selecting a character by focusing on it. The flashing row or column evokes a P300 ERP response in the EEG, allowing the classifier algorithm to determine the target row and column after averaging several responses, thus selecting the desired character. Although this research also uses a $ 6 \times 6 $ row/column flashboard design, the results extend to other flashboard types, such as the checkerboard \cite{townsend} format.

\begin{figure*}[!t]
\centering
\includegraphics[width=5.6in,height=3.5in]{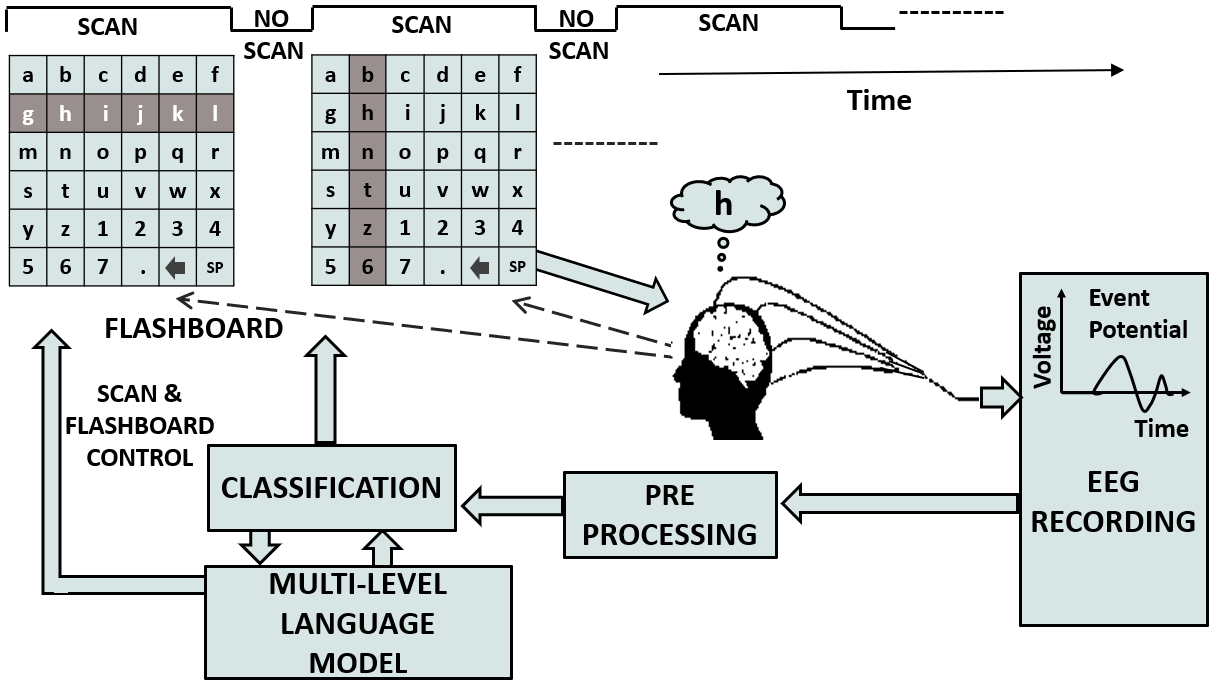}
\caption{A block diagram of the P300 speller based BCI used in 
this work. Subject EEG response to character highlights on a flashboard are shown.}
\label{blockdiag}
\end{figure*}

The speed of typing remains a significant limitation of the P300 speller (a block diagram of which is shown in Figure \ref{blockdiag}), along with the challenge of efficient classification across multiple subjects. This paper addresses both issues uniquely and comprehensively, using large language models for word prediction while also introducing novel {\it 'cross'} subject classifiers. Large language models, successfully applied in fields such as personalized learning \cite{Morris} and healthcare \cite{Lee}, have shown significant potential in BCIs to improve the speed and precision of user intent prediction \cite{Xu}.
These innovations in classifier training and language model integration, which draw on cross-disciplinary approaches, simplify the decoding of the EEG response and significantly enhance typing speed. Furthermore, this paper also demonstrates a novel methodology for seamlessly upgrading existing spellers with these new techniques without hardware changes. Most of the new techniques demonstrated in this paper are also applicable to other BCI evoked response methods.

Although designed as a communication medium, traditional signal classification methods in the P300 speller have not fully leveraged existing knowledge about the language domain and its associated correlations. As a simple example of such dependencies, note that it is more likely that the character {\it 'e'} is followed by {\it 's'} than {\it 'z'} in an English word \cite{mayzner}. Historically, while natural language has been used for years to improve classification in other areas such as speech recognition \cite{jelinek}, its use in the field of BCI \cite{speier1},\cite{speier2},\cite{nithin1} is a relatively recent movement. Although language models provide a means to exploit natural redundancies in speech \cite{speier_spell}, character selections in previous work have largely been treated as independent elements chosen from a set without prior information. Current random highlighting techniques, more specifically, ignore multi-word context \cite{kindermans1},\cite{kindermans2} along with word associations and word prediction given prior and partially formed words, thereby limiting communication efficiency. In fact, BCI studies using n-gram language models provide a poor representation of natural language as by disregarding context, they can potentially attach a high probability to character strings that do not formulate words. Therefore, in contrast to previous work, this research uses the additional contextual information available through increasingly sophisticated \cite{Kucera},\cite{texgen},\cite{roberta}-\cite{gptneo} language models to increase the efficiency of the BCI system.

The goal of this study is multifold, with a central theme of eliminating redundancy with sophisticated contextual word prediction algorithms,  allowing for efficient with and {\it “across”} subject classifier training. These classifiers minimize the need for subject-specific calibration, thereby facilitating the development of a single universal classification scheme. Performance of these models are compared to performance bounds with advanced large language models shown to be within $5 \% $ of the bound. Furthermore, this performance is shown to be far greater than that which can be achieved with character prediction.

Using extensive {\it offline} simulations based on multi-subject EEG data along with layered word completion
algorithms such as GPT2, BART, BERT in conjunction with Dijkstra's algorithm, the P300 speller performance is dramatically improved with both {\it 'within'} and {\it 'across'} subject classifiers. New flashboard and scanning
configurations are also developed to further optimize the speller performance and evaluate their relative merits. To overcome some limitations in previous work, smoothing algorithms are employed to enhance out-of-vocabulary (OOV) performance of word and character prediction.

The simulation techniques developed in this work allow for the comparison of many speller scenarios on large and representative targets. These include OOV words that would otherwise be infeasible in an online study due to the significant time requirements. An automatic next-stage enhancement of this research would then be to verify the projected gains with online studies. This will be the study of our future work.

The organization of this paper is as follows. In Section \ref{section_methods}, a detailed description is provided of the various techniques and algorithms used for the flashboard, stimuli, and classifiers, as well as character and word prediction. Section \ref{section_results} illustrates the results, while Section \ref{section_discussion} provides a summary of the results, which is followed by the section on conclusions and future directions.

\section{Methods}
\label{section_methods}
We will begin with a description of the adaptation of language models to the P300 speller.

\subsection{Language models}
In spoken English, vowels are well established to occur more frequently than consonants \cite{cornell}. Hence, from a timeline perspective of when they get highlighted in the flashboard scanning order, it is only natural that they occur earlier than some relatively rare consonants such as {\it 'x'} or {\it 'z'}. We will now describe how to perform such trade-offs using multi-level language models. 

\begin{figure*}[!t]
\centering
\includegraphics[width=5.7in]{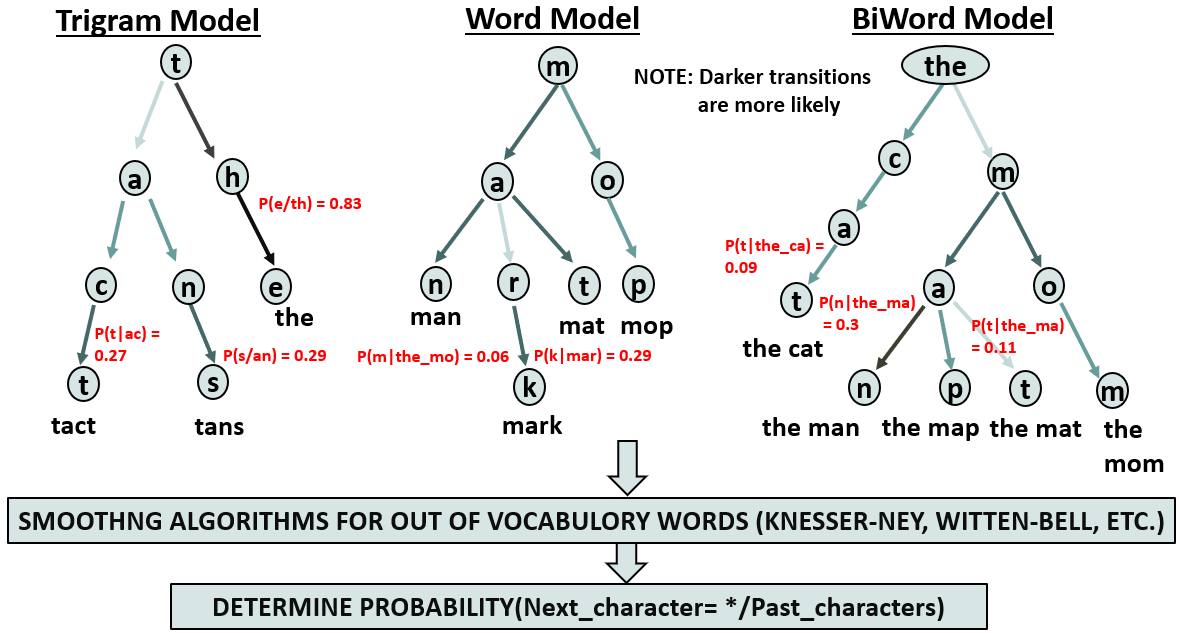}
\caption{Types of language models used. Starting with the simplest on the left to the most complex on the right}
\label{fig_langmodel}
\end{figure*}

Smoothing techniques \cite{jurafsky}, allow transitioning
across models when OOV words are encountered and
consequentially, complex models can be blended in with simpler
models. In Figure \ref{fig_langmodel}, from the simple {\it 'Trigram'} model to the most complex ({\it “Biword”}) 
which consists of pairs of consecutive words, an entire range
of models are depicted. As an example, consider the word
{\it 'cat'} and suppose {\it 'ca'} has already been spelled. 
Assume that {\it 'cat'} is not present in the language model. 
Hence, the probability of {\it 't' } given that {\it 'ca' } has
already been spelled (represented by the notation, $ p(\ulq t \urq|\ulq ca \urq) $ is zero. Smoothing evaluates this probability of OOV considering other contexts where {\it 'a' } and {\it ‘t’} have co-occurred through $ p(\ulq t \urq/\ulq a \urq) $. 
Multiple smoothing techniques were considered, and the Knesser-Ney smoothing (described in the following paragraph) was chosen
for its superior OOV performance \cite{kinderman3}. Also note
that, in general, the more sophisticated the language model, 
the conditional probability can be predicted with greater
precision \cite{openai}. Language models such as the one used
contain a large vocabulary, which results in high prediction
accuracy even as the character length increases.

Note that the convention adopted in the following terminology
is that $ Model(\ulq * \urq) $ denotes the frequency of occurrence of the characters $\ulq * \urq$. For example, {\it biword\_model('the quick')} is the frequency of the word {\it 'the quick'} in the biword language model. Also, as an example of conditional probability syntax, $ p\_biword (\ulq c \urq | \ulq the\,qui \urq) $ is the probability in the biword\-model of the character $ \ulq c \urq $ being next given that the partial phrase $ \ulq the\,qui \urq $ has been decoded.
\begin{multline}
p\_biword (\ulq c \urq | \ulq the\,qui \urq) = Max (biword\_model (\ulq the\,quic \urq) - d_1, 0) / biword\_model (\ulq the\,qui \urq)\\ +  d_1*L_1*p\_word(\ulq c \urq | \ulq qui \urq) 
\label{kn_eqn1}
\end{multline}
where $ L_1 $ is the normalizing constant equal to the number of distinct letters that follow $ \ulq the\,qui \urq $ in the biword model divided by the biword model count of $ \ulq the\,qui \urq$. If $ biword\_model (\ulq the\,qui \urq ) = 0, L_1 =1 $ to take care of out-of-vocabulary (OOV) conditions.
\begin{multline}
p\_word (\ulq c \urq | \ulq qui \urq) = Max(word\_model (\ulq quic\urq) - d_2, 0) /word\_model(\ulq qui \urq)\\ + d_2*L_2*p\_trigram (\ulq c \urq | \ulq qui \urq) 
\label{kn_eqn2}
\end{multline}
\begin{multline}
p\_trigram(\ulq c \urq | \ulq ui \urq)  = Max (trigram\_model (\ulq uic \urq) - d_3, 0) / trigram\_model(\ulq ui \urq) \\ + d_3*L_3*p\_bigram(\ulq c \urq | \ulq i \urq) 
\label{kn_eqn3}
\end{multline}
\begin{multline}
p\_bigram (\ulq c \urq | \ulq i \urq)= Max (bigram\_model (\ulq ic \urq) - d_4, 0) / bigram\_model(\ulq i \urq) \\ + d_4*L_4*p\_unigram(\ulq c \urq) 
\label{kn_eqn4}
\end{multline}
\begin{dmath}
p\_unigram (\ulq c \urq) = unigram\_model (\ulq c \urq) / 
unigram\_model (\ulq \urq) 
\label{kn_eqn5}
\end{dmath}
where $ unigram\_model(\ulq \urq) $ = {\it total\,model\,entries} and $ d_1, d_2, d_3, d_4 $ are tunable parameters with $ 0 <=  d_1, d_2, d_3, d_4 <=  1$. All these {\it d} values were set to the commonly used midrange default of 0.5. Just as $ L_1$ is a biword model normalizing constant in equation \ref{kn_eqn1} whose derivation was described earlier with an example, $ L_2, L_3, L_4 $ are also normalizing constants similarly obtained from the corresponding lower-order language models. More specifically, in the example used, $ L_2 $ is the normalizing constant equal to the number of distinct letters that follow $ \ulq qui \urq $ in the word model divided by the word model count of $ \ulq qui \urq $. Furthermore, for satisfying the OOV conditions, if $ word\_model (\ulq qui \urq) = 0, L_2 =1 $. If $ trigram\_model (\ulq ui \urq) = 0, L_3 = 1 $ and if $ bigram\_model (\ulq i \urq) = 0, L_4 = 1 $ thereby accounting for the values of $ L_1 $...$L_4 $ in all OOV cases.  
In summary, through equations \ref{kn_eqn1}-\ref{kn_eqn5}, the most complex model in Figure \ref{fig_langmodel} falls back to lower-level models with OOV characters.

\subsection{Data Collection}
\label{subsection_data_collection}
\begin{figure}[!t]
\centering
\includegraphics[width=3.2in,height=1.6in]{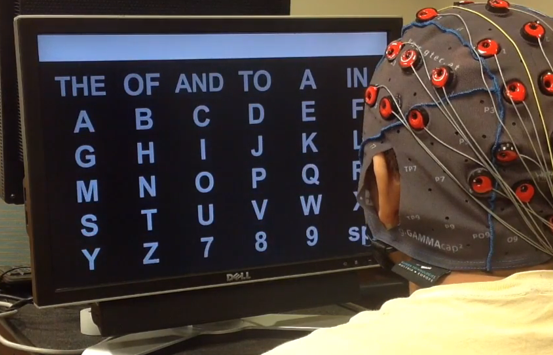}
\caption{BCI display and volunteer EEG recording during stimulus presentation.}
\label{fig_volunteer}
\end{figure}
Data for offline analyzes were obtained from 78 healthy volunteers with normal or corrected normal vision between the ages of 20 and 35. Figure \ref{fig_volunteer} shows the BCI display flashing before a volunteer whose response is being recorded. All data were acquired using GTEC amplifiers, active EEG electrodes and electrode cap (Guger Technologies, Graz, Austria); sampled at {\it 256\,Hz}, referenced to the left ear; grounded to {\it AF_Z}; and filtered using a bandpass of {\it 0.1 – 60\,Hz}. The set of electrodes consisted of 32 channels placed according to a previously published configuration \cite{jessicalu}. We used a window of 600 ms at 256 Hz, resulting in 154 time points per channel. We then downsampled the signal by a factor of 12 to obtain a feature vector that was 32×13=416 features. The system also used a {\it 6×6} character grid, row and column flashes, and a stimulus duration of 100\,msec and an interval between stimulus of 25\,ms for a stimulus onset asynchrony of 125\,msec. The channels used were {\it Fpz, Fz, FC1, FCz, FC2, FC4, FC6, C4, C6, CP4, CP6, FC3, FC5, C3, C5, CP3, CP5, CP1, P1, Cz, CPz, Pz, POz, CP2, P2, PO7, PO3, O1, Oz, O2, PO4, PO8} again based on a published montage \cite{jessicalu}. 

\subsection{Feature Extraction: Training Methodology}
\label{subsection_feature_extraction}
The stepwise linear discriminant analysis (SWLDA) as described by Speier \cite{speier2} et al. and originally used in \cite{farwell} was used as a classifier. SWLDA is a stepwise method that sequentially adds or removes features based on their statistical significance, optimizing the model for better classification. It was chosen for its efficiency in handling high-dimensional data and its proven effectiveness in BCI applications \cite{sellers}.

SWLDA utilizes a discriminant function, which is determined in the training step \cite{draper}. The SWLDA classifier was trained using one of two methods:  Across-subject cross validation (ASCV) (also known as leave-one-subject-out cross  validation) and within-subject cross validation (WSCV). Although both use SWLDA, note that these two techniques fundamentally differ in how entire subject responses are split between what was used for training and what was used for testing. \\
\textbf{a) ASCV:} Out of $ N $ subjects, one subject was chosen to be the test subject, while the data of the other $ N-{\it 1 }$ subjects were used to train the SWLDA classifier. The SWLDA classifier then attempted to take the data of the test subject and spell out the target phrase. Note that this technique finally produces one classifier table for all $ N $ subjects. \\
\textbf{ b) WSCV:} In the WSCV, a two-fold cross-validation was performed using SWLDA on the EEG data of each individual subject (unlike ASCV), resulting in a unique classifier output for each subject. More specifically, the output phrase contained at least $ 20 $ characters and had 120 flashes for each character that was split into three sets. During the first test iteration, the first set was chosen as the test phrase, while the other two sets were used to train the SWLDA classifier. On completion, the same procedure would repeat, but with the second and third sets being the test phrases during the second and third trial, respectively. This method would then be replicated on all subjects.

During training, class labels were predicted using ordinary least squares. Next, the forward step-wise analysis adds the most important features, and the backward analysis step determines and removes the least important features. This process repeats until the number of features remains constant after a set number of iterations or until the number of required features is met. The final selected features are then added to the discriminant function. The flash score, $ i $ for the character $ t $, $ y_t^i $ can then be calculated as the dot product of the feature weight vector with the features of the signal from that trial.

\subsection{Virtual flashboard design using a language model}
\begin{figure}[!t]
\centering
\includegraphics[width=5.6in,height=2.4in]{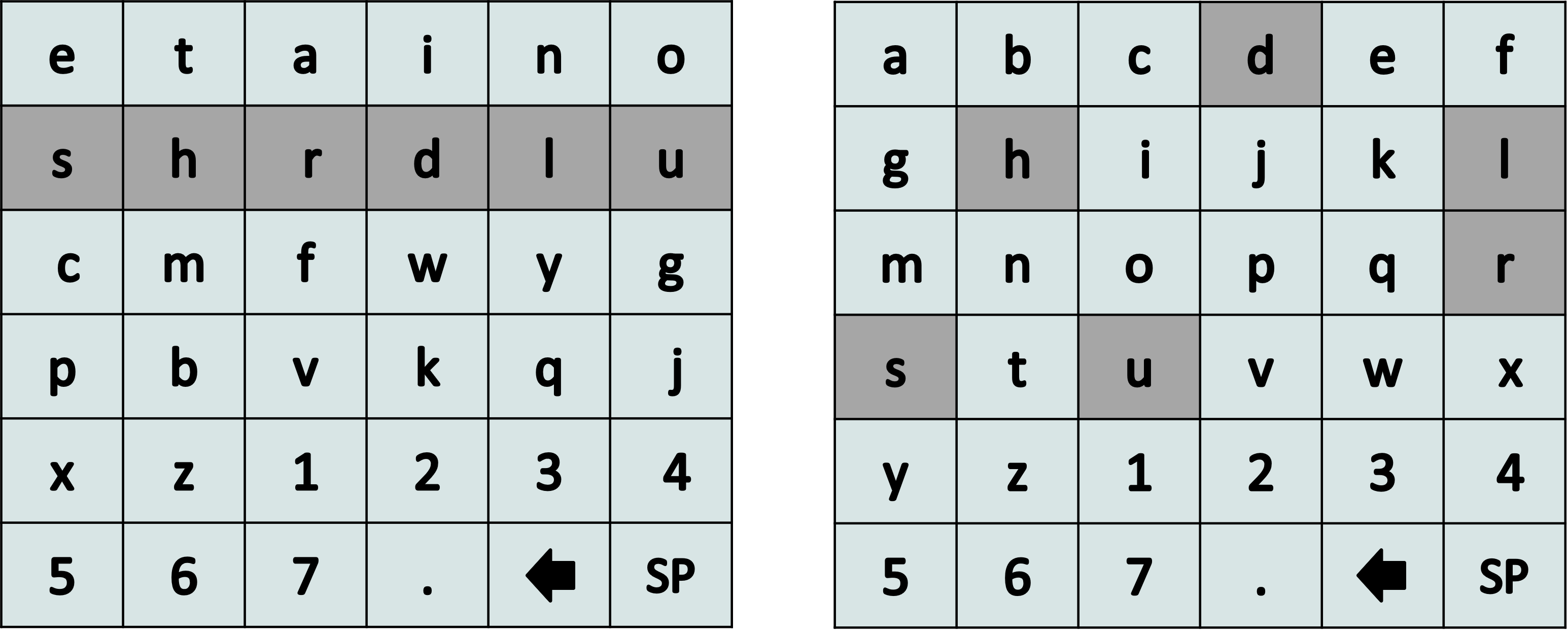}
\caption*{({\it a})     \hspace{7cm}   ({\it b})}
\caption{a) Probabilistic flashboard highlighting order (Prob({\it Next character/Earlier spelled characters}) in a sequential frequency-sorted flashboard. {\it SP} in the flashboard denotes space between words\\ b) Sequential flashboard characters shown in Figure \ref{fig_seq}a are {\it virtually} mapped onto a conventional alphabetical flashboard}
\label{fig_seq}
\end{figure}

\begin{figure}[!t]
\centering
\includegraphics[width=5.6in,height=2.4in]{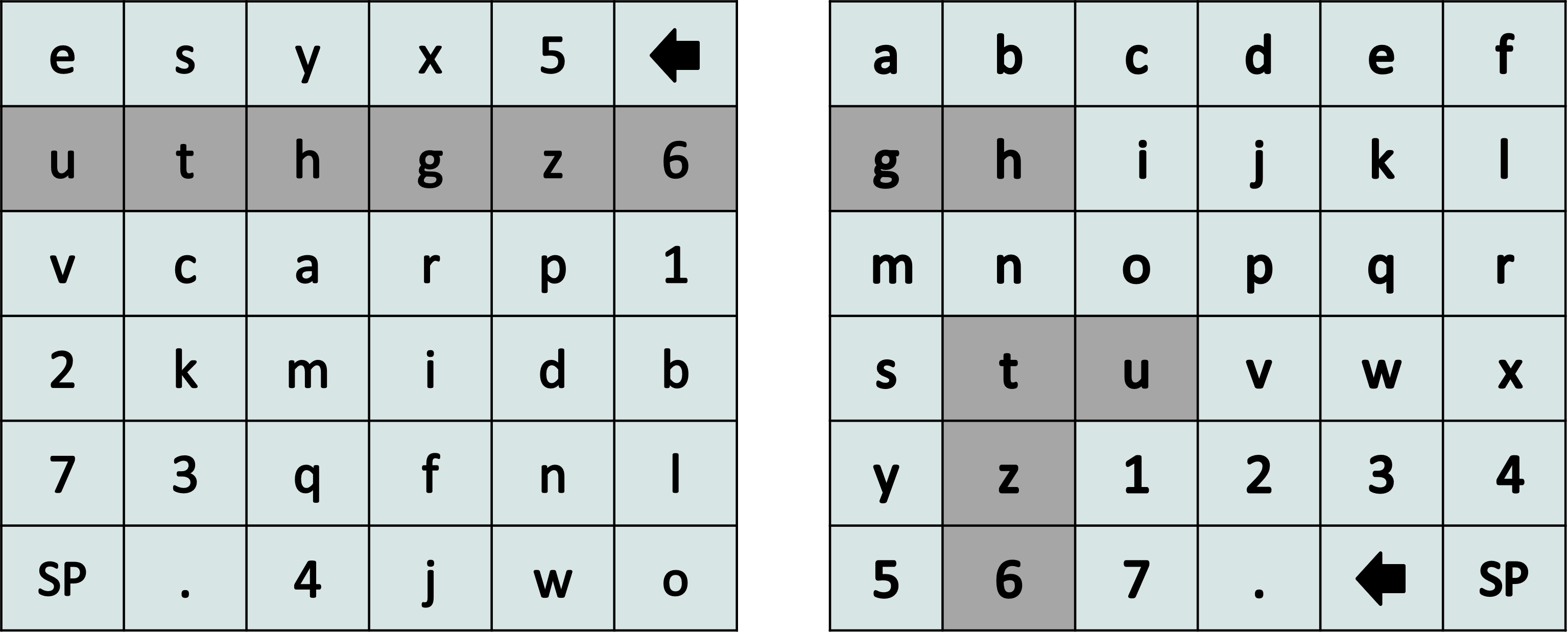}
\caption*{({\it a})     \hspace{7cm}   ({\it b})}
\caption{(a) Highlighting in a diagonal flashboard design where the probabilities represent the Prob({\it Next character/Earlier spelled characters}). Most likely characters {\it e, t, a, i ...} are organized along diagonals. Note that {\it SP} in the flashboard denotes space between words\\
b) Diagonal highlighted flashboard characters in Figure \ref{fig_diag}a are {\it virtually} mapped onto a conventional alphabetical flashboard}
\label{fig_diag}
\end{figure}

The flashboard is a central part of a P300-based BCI system, as seen earlier in Figure \ref{blockdiag}. In this section, we will describe schemes that organize the highlighted flashboard characters based on their probability of occurrence. Note that these schemes are mapped {\it virtually} onto conventional static flashboards, whereby only the highlighted set of characters is modified, and not the underlying flashboard itself. In other words, the flashboard is constant through the flashing sequence, though the highlighted characters can be arbitrarily chosen. By doing so, this readily overlays onto any flashboard in current use, including popularly used random highlighting. In related work, Townsend \cite{townsend} proposes the checkerboard paradigm where flashing occurs in non-adjacent groups as opposed to row-column to reduce errors. \\
\textbf{a) Sequential flashboard:} The flashboard highlighting is {\it virtually}  modified (as shown in Figure \ref{fig_seq}) so that each set of highlighted characters is picked in a frequency weighted rather than alphabetically. More specifically, higher frequency characters are highlighted earlier, whereby the initial flashes are more likely to select the target character. The results of these modifications will be described in Section \ref{section_results} and discussed in Section \ref{section_discussion}. \\
\textbf{b) Diagonal flashboard:} Consider the diagonal design in Figure \ref{fig_diag}, where the most likely highlighted characters are organized along diagonals. In the figure, note that the most likely letters \textit{e, t, a, i, n, o} are placed on the main diagonal, followed by the next likely letters on the upper and lower diagonals. This design forces characters that appear in similar contexts to flash separately, making them easier to distinguish. Additionally, this \textit{virtual} technique can easily overlay onto static alphabetically arranged flashboards. Gains from these schemes are provided in Section \ref{section_results}.
\subsection{Further optimization of the scanning order}
This section describes modifications to adapt the P300 BCI scanning to optimize performance.\\
\textbf{a) Weighted scanning order:} Conventional random and deterministic flashing are straightforward schemes, where the flashing set of characters is picked deterministically or round-robin. In contrast, in the proposed weighted flashing enhancement to conventional flashing, the highlighted set is picked based on that set's probability as opposed to the random scheme where the sets are assumed to be equi-probable. The idea is that the more likely characters are highlighted earlier.\\
\textbf{b) Dynamic stopping:} Once a character is determined at the decoder, continuing the flashing creates an unnecessary reduction in speed, which is avoided by using a {\it“dynamic stopping”} strategy. Flashing of the current character is terminated and the next character flash is begun.
\begin{figure*}
\centering
\includegraphics[width=6.1in,height=2.9in]{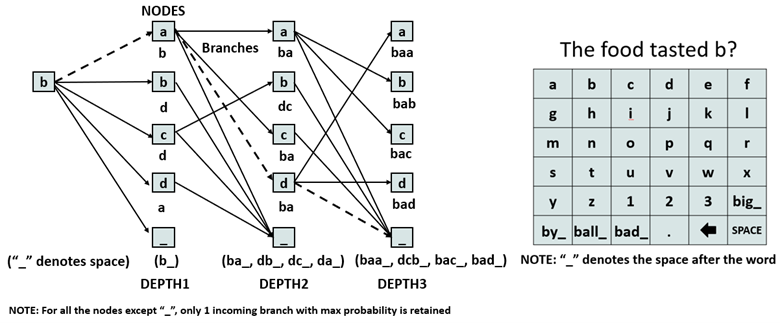}
\caption*{\hspace{2cm}({\it a})     \hspace{8cm}   ({\it b})}
\caption{(a) Implementation of Dijkstra's algorithm to find most likely word completions. (b) Corresponding flashboard showing final most likely word completions with 4 characters have been replaced by the word choices. Note: “\textunderscore” represents space between words.}
\label{dijkstra_diag}
\end{figure*} 

\begin{figure}
\centering
\includegraphics[width=2.5in,height=2.4in]{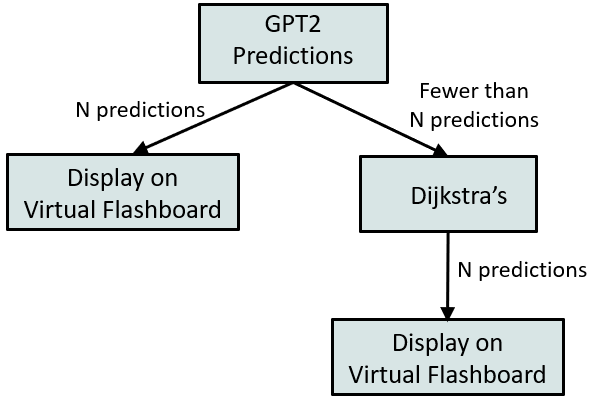}
\hfill
\caption{Layered GPT2 word choice selection to obtain most 
likely {\it N } word completions with fallback to Dijkstra's 
algorithm}
\label{fig_layered_gpt2}
\end{figure}
\subsection{Word prediction algorithms and methodology}
\label{methodology}
Similarly to word completion suggestions that are available when messaging using smartphones, flashboard efficiency can be
likewise increased by replacing unlikely characters with word
choices. This section describes algorithms to compute these word choices.\\
\textbf{a) Dijkstra’s Algorithm:} The classic technique for 
traversing a tree or a graph is either a depth-first search or 
a breadth-first search, which are well known algorithms \cite{efreq}. Although speedwise efficient, a key
limitation is that it does not provide a technique for 
evaluating OOV words, which occur when the partially spelled 
word is absent from the language model. In contrast, our approach is to use Dijkstra’s algorithm \cite{Dijkstra} for word prediction. This is implemented in the form of a trellis similar to dynamic programming techniques as shown in Figure \ref{dijkstra_diag}. As shown in the figure, every node in a trellis stage has $ M $ incoming possibilities (where $ M $ is the number of unique characters on the flashboard) out of which only the one with the highest probability is retained. In doing so, the complexity is reduced, as every possibility does not have to be exhaustively evaluated. Further, as more
stages are incorporated into the trellis, longer length word choices are obtained, resulting in higher evaluation time and complexity. For each branch at a given stage, probabilities are determined by using the frequency count of branch characters given prior characters using the language model along with smoothing. At every stage, the highest probability completions then provide the most likely word completions.\\

\textbf{b) Large Language Models:}  Numerous popular language models were used in this research, such as Roberta \cite{roberta}, BERT \cite{bert}, XLnet \cite{xlnet}, BART \cite{bart}, GPTNeo \cite{gptneo} and GPT2 \cite{openai},\cite{transformer}. The description provided here will be minimal as the literature on them is easily available besides being relatively simple to obtain software implementations. One particular note on GPT2, as it is newer, advanced versions are quoted in the press quite often. A large unsupervised transformer \cite{openai},\cite{transformer} based model released by {\it “OpenAI”}, GPT2 is an enhanced deep neural network enhancement to the earlier GPT model and is powerful enough to generate coherent passages of text. In contrast to earlier models, GPT2 uses 'attention mechanisms' to focus on the segments of the input text that it believes are the most relevant. In the context of BCI, it can become a sophisticated tool that predicts the subsequent word in a partially formed sentence given prior words. Although GPT2 is powerful, it does not have the ability to predict OOV. However, it can be combined with Dijkstra’s algorithm (as shown in Figure \ref{fig_layered_gpt2}) to account for OOV words, thus creating a layered approach similar in spirit to character-based smoothing algorithms. More specifically, as seen in the figure, in case the GPT2 word choices result in an empty or a partially complete set of choices, one can revert to Dijkstra’s to predict OOV words.

\textbf{c) Performance Bounds:} Both character and word performance bounds are obtained by providing a high probability for the correct next character/word. More specifically, let us first consider the character performance bound. Suppose that the word to be spelled is {\it 'cat'} and suppose that {\it 'ca'} has already been spelled. Then we can provide a high probability for the character {\it 't'}. Also note that all the other characters are uniformly set to small values (since the total probability has to sum up to 1). Similarly, for the word bound, we can set a high probability for the correct next word while any other word choices are given a 0 probability. In our bound computation, the correct word and character probabilities are set to 0.5. Note then that the performance for an individual depends on the individual P300 signal strength.

\subsection{Decoder}
In this paper, a simple decoding scheme based on thresholding is used to deduce the character or word. First, the score for each flash is computed by, $ y_t^i  = {\mathbf w}{\bf .z_t^i }$ where the feature weight vector is determined by feature extraction as described in the previous subsection. With the assumption that the distributions were Gaussian \cite{speier2}, the {\it 'attended'} and {\it 'non-attended'} signals (note that this terminology refers to highlighted and non-highlighted characters on the flashboard) were found and given by the authors. 
\begin{dmath}
 f(y_{t}^{i}/x_{t}) = \left\{
    \begin{array}{ll}
         \frac{1}{\sqrt{2\pi\sigma_{a}^{2}}} 
   e^{\frac{(y_{t}^{i} - \mu_{a})^2}{2\sigma_{a}^2}} & \mbox{if } x_{t} \in {\mathbf A_{t}^i}\\
   \frac{1}{\sqrt{2\pi\sigma_{n}^{2}}} 
   e^{\frac{(y_{t}^{i} - \mu_{n})^2}{2\sigma_{n}^2}} & \mbox{if } x_{t} \notin {\mathbf A_{t}^i}\\     
    \end{array}
\right.
\label{eqnPdf}
\end{dmath}    
where $ {\mathbf A_{t}^i} $ is the set of characters illuminated for the $ i^{th} $ flash for the character $ t $ in the sequence. $ \mu_a,\sigma_a^2,\mu_n,\sigma_n^2 $ are the means and variances of the distributions for the attended and non-attended flashes, respectively. 
\begin{dmath}
 P (x_t|{\bf y_t},x_{t-1},...,x_0  )= 
 \frac{P (x_t | x_{t-1},...,x_0)} {P ({\bf y_t} | x_{t-1},..,x_0  )).P ({\bf y_t} |x_{t},..,x_0  )}
 \end{dmath}
 \begin{dmath}
 =\frac{1}{Z}.P (x_t | x_{t-1},..,x_0  )\prod_if(y_t^i | x_t)
\end{dmath}
where $ P(x_t|x_{t-1},.., x_0) $ is the prior probability of a character given the history, $ f(y_t^i|x_t) $ are the pdf’s from equation \ref{eqnPdf}, and $ Z $ is a normalizing constant. The language model is used to derive the prior as per
\begin{dmath}
P(x_t | x_{t-1},..,x_0  )=  Count(x_t,x_{t-1},..,x_0  ) )/\sum_{x_t}Count(x_{t},x_{t-1},..,x_{0})
\end{dmath}
where \( Count(x_{t},x_{t-1},..,x_0) \) is
the number of occurrences of the string \begin{math} x_{t}x_{t-1}..x_{0} \end{math} in the corpus.
A threshold probability, $ P_{Thresh} $, is then set to determine when a decision should be made. The program flashes characters until either $ max_{x_t} \,P (x_t|{\bf y_t}, x_{t-1},.., x_0) \geq P_{Thresh} $ or the number of flash sets reaches a predetermined maximum value. The classifier then selects the character that satisfies $ arg\, max_{x_t}\, P (x_t|{\bf y_t}, x_{t-1},.., x_0) $.

The subject can correct the typing errors by invoking the backspace provided in the flashboard. As a result, all the typing errors are fixed, reducing the overall error rate to zero. Alternatively, one can have an autocorrect wherein typed words are checked for presence in a dictionary and erred words are replaced by closest most likely words. This technique is faster, but some mistyped words are likely to either be mis-corrected or left uncorrected, which would result in a non-zero final error rate. Although a vast array of more powerful decoding schemes can be used that involve hidden Markov models, machine learning, and particle filters \cite{speier2}, note that these enhancements only complement the techniques introduced in this paper and therefore do not deter the results. 

\subsection{Simulation overview}
\begin{figure*}[!t]
\centering
\includegraphics[width=5.5in,height=1.7in]{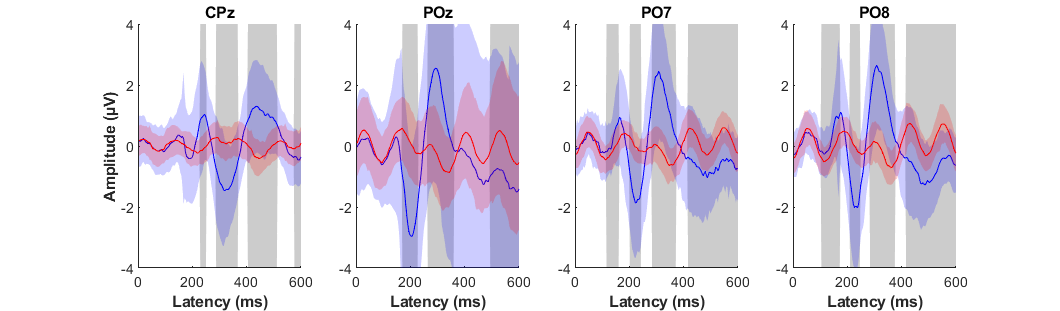}
\caption{Waveform analysis across all subjects in the dataset, subjects for four channels. These are the average ERPs for target (blue) and non-target (red) stimuli, which are averaged across all dataset subjects. Shaded regions represent the standard error of the mean, and vertical gray regions represent significant differences after using false discovery rate to account for multiple comparisons.}
\label{P300sig}
\end{figure*}
Across all subjects in the dataset, the average waveforms found for the attended and non-attended stimulus responses are shown in Figure \ref{P300sig}. Using the EEG signal for the online sessions, the stimulus responses were grouped according to whether they corresponded to the character the subject was currently trying to spell. For each subject, the average response for attended and non-attended stimuli was produced. Then a global average was produced by averaging these signals across subjects. 
For this discussion, we focus on four channels that we previously determined to be the most influential in our P300 analyzes \cite{speierP300}. In each of these channels, a defined negative peak was observed around 200 ms after attended stimuli, followed by a positive peak around 300 ms. The peaks in the Cz channel were slightly later than the peaks in the parieto-occipital channels. In each of these channels, the difference between the attended and non-attended signals was significant in each of these peaks. In addition, there were smaller peaks in both attended and non-attended signals with an interval of 125 ms, which reflects the stimulus onset asynchrony. These peaks likely reflect attended stimuli before or after the current stimulus due to the overlapping response intervals. 
Statistical variation from such small data sets would limit the confidence in the simulation results. As a result, a large and meaningful data set is used as simulation data. The text of {\it 'Declaration of Independence'} (DOI) is used as the simulation data \cite{doi}. The full length of this document is 7,892 characters after removing all punctuation and special characters. The DOI was chosen as the simulation dataset as it is concise, well-known, and contains words of varying length, including a number of out-of-vocabulary (OOV) words that stress the simulation. Furthermore, to keep the simulation accurate, numerous subject responses are used to provide a wide range of brain activity. 
 
During simulations with the DOI dataset, the attended and non-attended (once again, this terminology refers to whether the desired character is present or absent in the highlighted flashboard character) scores for a given subject were chosen from the laboratory results for that particular subject. The laboratory results are sampled and subsequently modeled by a two-state Markov chain, which models the attended and the non-attended states. Thus, there are four possible sampling combinations depending on whether the current and previous state was attended or non-attended. When a character or word is decoded, it is compared to the intended character or word. In the event of an error, a {\it Backspace} is initiated as the subsequent character. Following this, the deleted character/word is flashed again. The net effect of this is to {\it undo} the errors, thereby reducing the final error rate to zero. However, note that this could possibly require multiple attempts and a long follow-up time. The follow-up time could be significantly restricted in addition to tolerating residual error, and this evaluation could be a potential future study.
\subsection{Performance evaluation metrics}
\label{itr_section}
Information Transfer Rate (ITR) is a crucial metric in the evaluation of Brain-Computer Interfaces (BCIs), quantifying the speed and precision of communication \cite{mcfarland1}. The metric is attractive for several reasons: It is derived from the principles of information theory, it combines the competing statistics of speed and accuracy, and it reduces performance to an information transfer problem that can be compared between applications \cite{pierce}.
It is typically calculated using the formula: $ \frac{log_{2}(N+1)}{T+P_{f}(T_r+T_c)} \times 60 \; bits/minute $ where $ N $ is the number of characters in the speller (grid size), $ T $ is the total time taken to complete the task (including pauses, corrections, etc.). Note that $ P_f $ is the probability of incorrectly selecting a character, $ T_r $ is the time required for a single character selection, and $ T_c $ is the time taken for error corrections.

The second metric used is the retry rate (or error rate), which measures the number of backspaces initiated as a fraction of the overall communicated characters. Note that when a character or word is decoded in error, the next target character is assumed to be a backspace. The objective is to attempt to mimic a subject by correcting the misinterpreted error or word. This process repeats until the character/word is decoded correctly. Note that the accuracy will reduce to zero only if the error rate is sufficiently low. It is possible for the system to generate errors faster than the user can correct them. However, in our analysis, in almost all cases, we see high enough accuracies that users can make all corrections as they happen, resulting in a perfect output. In the case where simulation of a particular character or word requires more than 75 flashboard scans, it is abandoned, and an ITR of 0 (100\% error rate) is assigned for that particular character or word and the simulation moves on to the next character or word. 

Papers in this area \cite{gogna},\cite{kshirsagar} have used sensitivity and specificity as performance metrics. However, metrics such as sensitivity and specificity measure the frequency of correct letters in a subject's output. In the current study, all errors are manually corrected, so no errors remain in the output text. Therefore, metrics such as sensitivity and specificity are 100\% for all subjects by necessity. 

Some errors (e.g., obvious typos) do not affect the readability of text and could therefore be allowed to remain in the final output without affecting the meaning. Allowing these errors to remain could then potentially result in a more efficient system because it would not waste time forcing the user to make these corrections. Language models could also be used to automatically correct some of these errors  \cite{speier2},\cite{speier3} if they are not corrected manually. 
\begin{figure*}
\begin{subfigure}[b]{0.5\columnwidth}
\centering
\includegraphics[width=2.8in,height=2.8in]{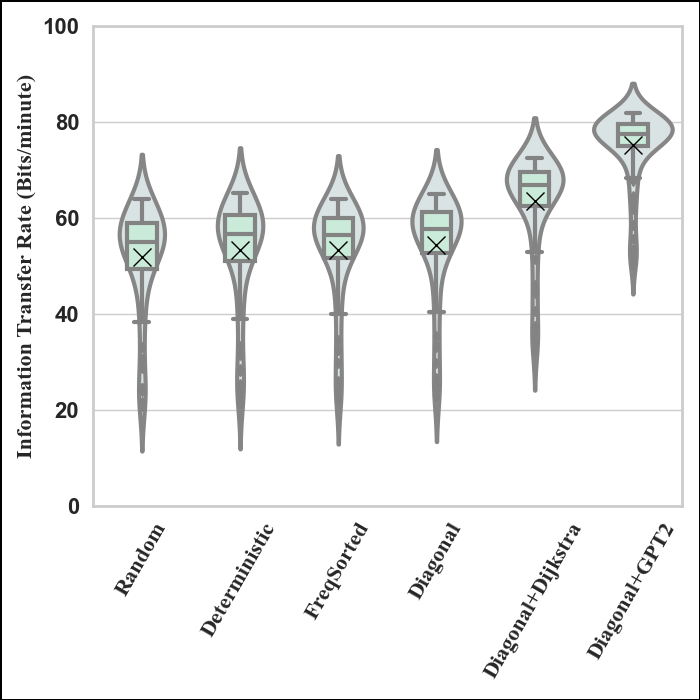}
\end{subfigure}%
\begin{subfigure}[b]{0.5\columnwidth}
\centering
\includegraphics[width=2.8in,height=2.8in]{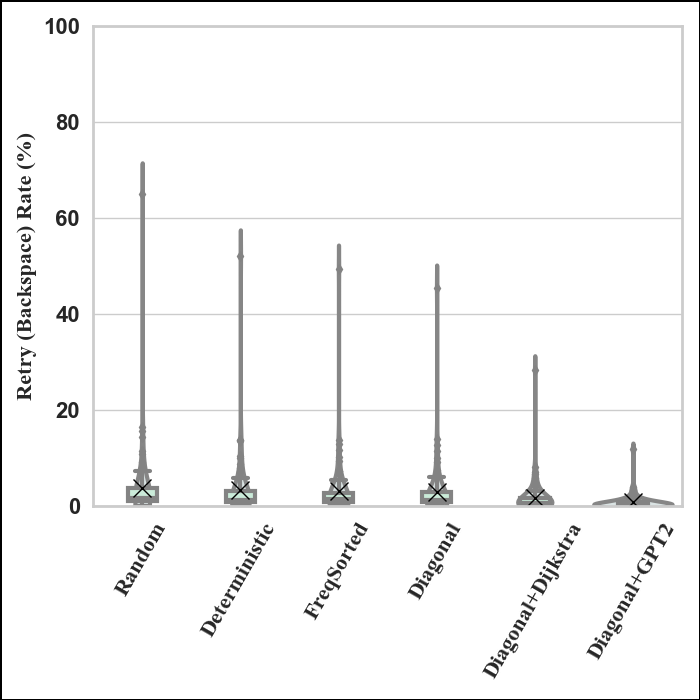}
\end{subfigure}
\caption{a) ITR performance comparison of different schemes for the {\it “Within Subjects”} (WSCV) training b) Error rate \\
\textbf{NOTE:} Violin plot containing an embedded box-and-whiskers that shows the minimum, first quartile, median, third quartile, and maximum data. {\it ‘X’} is the mean ITR.}
\label{wscv_only}
\end{figure*}
Both the Shapiro-Wilk test \cite{nonparametric} and the Kolmogorov-Smirnov test \cite{nonparametric} on the results comprehensively reject normality $( p < 10^{-3} )$. Consequently, we use nonparametric tests for the hypothesis. \begin{figure}
\begin{subfigure}[b]{0.49\columnwidth}
\centering
\includegraphics[width=2.77in,height=2.8in]{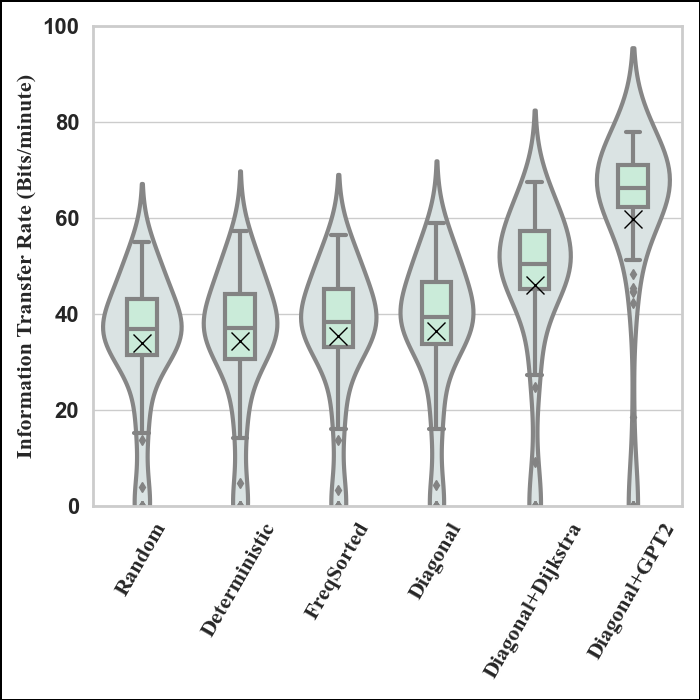}
\caption{}
\end{subfigure}
\begin{subfigure}[b]{0.5\columnwidth}
\centering
\includegraphics[width=2.8in,height=2.8in]{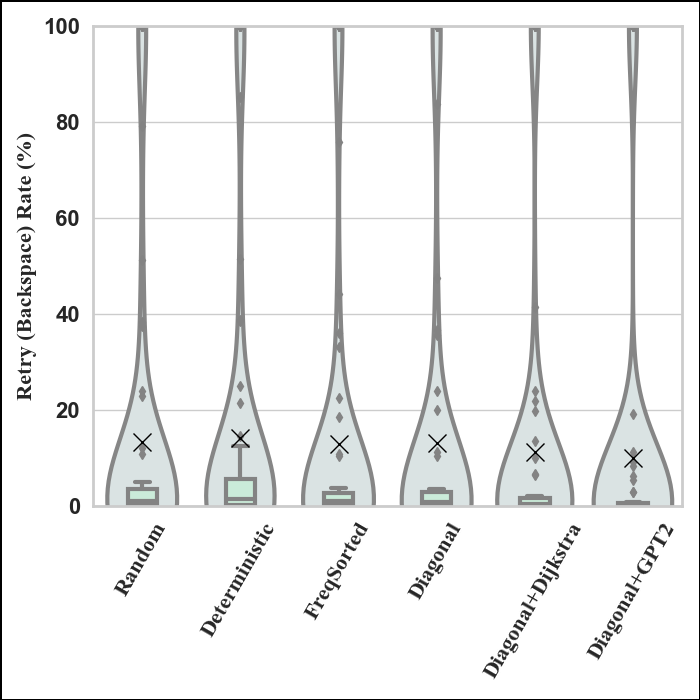}
\caption{}
\end{subfigure}
\caption{a) ITR performance comparison of different schemes for the {\it “Across Subjects”} (ASCV) training b) Error rate\\
}
\label{ascv_only}
\end{figure}
testing. The Kruskal-Wallis test \cite{nonparametric} was performed on the data, $( p < 10^{-6} )$ rejecting the null hypothesis of each scheme's ITR, error rate originating from
the same distribution. Hence, the Wilcoxon signed rank test \cite{nonparametric} for paired samples is then conducted to determine the statistical significance.
\section{Results}
\label{section_results}
This section details the performance evaluation of various schemes.
\begin{table}[ht!]
\large
\centering
\begin{tabular}{|c | c  | c|} 
 \hline
 \multicolumn{1}{|c|}{\multirow{2}{*}{\textbf{ Scheme}}} & 
 \multicolumn{1}{c|}{\textbf{ WSCV ITR}} &
  \multicolumn{1}{c|}{\textbf{ ASCV ITR}}\\
   & (bits/min)	& (bits/min) \\ 
 \hline 
 \hline
Random	          & $ 51.93 \pm 11.2 $	&   $ 33.95  \pm 14.4  $  \\ [0.25ex]\hline
Deterministic     &	$ 53.42 \pm 11.4 $          &	$ 34.49  \pm 14.9       $\\ [0.25ex]\hline
FreqSorted        &	$ 53.44 \pm 10.8 $          &	$ 35.44  \pm 15.0     $\\ [0.25ex]\hline
Diagonal          &	$ 54.43 \pm 11.0 $          &	$ 36.51  \pm 15.4     $\\ [0.25ex]\hline
Char Bound        &	$ 55.86 \pm 10.5 $          &	$ 37.97  \pm 14.6     $\\ [0.25ex]\hline
Dijkstra &	$ 63.59 \pm 10.1 $          &	$ 46.05  \pm 17.8    $\\ [0.25ex]\hline
Roberta &	$ 59.01 \pm 10.6 $          &	$ 41.19  \pm 16.6    $\\ [0.25ex]\hline
Bert &	$ 62.81 \pm 10.1 $          &	$ 45.25  \pm 17.6    $\\ [0.25ex]\hline
XLnet &	$ 65.51 \pm 9.8 $          &	$ 48.18  \pm 18.4   $\\ [0.25ex]\hline
Bart &	$ 71.64 \pm 8.6 $          &	$ 55.23  \pm 20.1    $\\ [0.25ex]\hline
GptNeo &	$ 75.30 \pm 7.5 $          &	$ 59.66  \pm 21.0    $\\ [0.25ex]\hline
GPT2     &	$ {\bf 75.31 \pm 7.5} $          &	$ {\bf 59.75 \pm 21.0}  $\\ [0.25ex]\hline
Word Bound     &	$ 78.21 \pm 6.4 $          &	$ 62.84 \pm 21.8  $\\ [0.25ex]\hline
\end{tabular}
\caption{Information transfer rate (mean $\pm$ standard deviation format) in bits/minute of various schemes for WSCV and ASCV subject training. The scheme with the highest mean ITR is shown in bold.}
\label{table1}
\end{table}
Figures \ref{wscv_only}a and \ref{wscv_only}b shows the performance of different schemes for the $ 78 $ subjects along with their error rates in a {\it violin} plot along with an embedded traditional box and whiskers plot. The violin plot depicts the density of the data in the form of a histogram, while the embedded box and whiskers plot provides relevant statistics. In this case, the classifier was trained on each subject individually ({\it WSCV} training of subsection \ref{subsection_feature_extraction}). A maximum of six word suggestions were provided in this simulation, with numbers in the flashboard replaced with the word suggestions. In Figure \ref{wscv_only}b, the error rate captures the backspace rate due to a wrong character decoding. However, note that the final error rate in the output text is zero in all these cases. Further, ITR was calculated as defined in section \ref{itr_section}.

 Using GPT2 word completion was the configuration that provided the highest average WSCV ITR (75.31 bits/minute), which was significantly higher than all other configurations other than GPTNeo (75.30 bits / minute, p = 0.01). In general, the standard deviation obtained using WSCV schemes is much lower compared to ASCV training. For example, while using GPT2, the standard deviation of WSCV was 7.5 bits/minute, which is almost a third lower than the equivalent ASCV number of 21. The width of the violin plots also shows the standard deviation in the results through the width of the histogram for each scheme. In ASCV training, the same trend holds with the GPT2 configuration providing the highest ITR (59.75 bits / minute), as seen in Figure \ref{ascv_only} and Table \ref{table1}. Once again, note that Figure \ref{ascv_only} shows the same results as Figure \ref{wscv_only}, but for the ASCV training of Section \ref{subsection_feature_extraction}. 

 In both Figures \ref{wscv_only}, \ref{ascv_only} and in Table \ref{table1}, the character and word performance bounds computed for each subject (as described in Section \ref{methodology}) are shown. These limits are a means of studying the efficacy of each particular scheme and its relative performance compared to the best achievable. The performance of the diagonal scheme is within $ 2\% $ of the character bound as seen in Table \ref{table1} (54.43 vs. 55.86 bits / minute for WSCV and 36.51 vs. 37.97 bits / minute for ASCV). Following similar trends, regardless of the WSCV/ASCV training methodology, GPT2/GPTNeo achieves performance within 4\% of the word bound (75.31 vs 78.21 bits/minute for WSCV and 59.75 vs 62.84 bits/minute for ASCV). 
 
\begin{figure*}
\begin{subfigure}[b]{0.5\columnwidth}
\centering
\includegraphics[width=2.8in,height=2.8in]{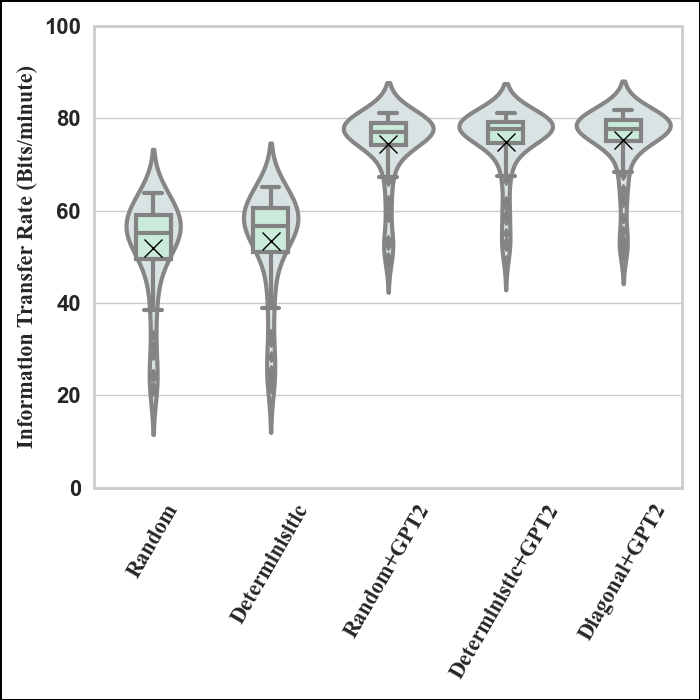}
\caption{Within Subjects}
\end{subfigure}%
\begin{subfigure}[b]{0.5\columnwidth}
\centering
\includegraphics[width=2.8in,height=2.8in]{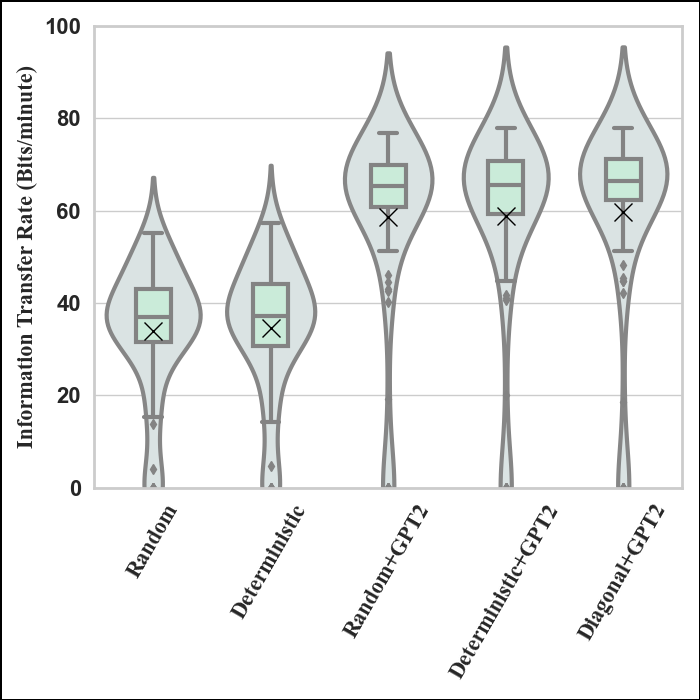}
\caption{Across Subjects}
\end{subfigure}
\caption{Violin plot of the information transfer rate of conventional flashboards, showing the effect of word completion on performance improvement for {\it a) “Within Subjects”} (WSCV) and {\it b) “Across Subjects”} (ASCV). \\ \textbf{NOTE:}{\it “Regular”} refers to the traditional random scan with alphabetical flashboard while {\it “Regular-Wcomp”} refers to the same but with word suggestions included}
\label{wscv_ascv}
\end{figure*}

Figure \ref{wscv_ascv} shows the performance of traditional schemes when augmented with good word completion and prediction choices. As shown in Figure \ref{fig_seq}, using the concept of virtual flashboard, word completion schemes become a simple overlay on conventional flashboards. Hence, one then observes the performance gains achieved by using this word completion retrofit technique. 
\section{Discussion}
\label{section_discussion} 
This study utilized a large set of subject EEG responses to simulate and optimize several novel flashboard highlighting strategies. These simulations were conducted on a standard HP laptop equipped with an Intel i7 processor and 16 GB of memory. Furthermore, the ITR was calculated as defined in Section \ref{itr_section}.

In the within-subject (WSCV) analysis, the frequency-sorted and diagonal schemes outperform traditional ones. For example, the mean ITR for the diagonal scheme was $54.4$ bits/minute, showing a modest improvement over the random scheme ($51.9$ bits/minute, $p < 0.001$). The diagonal scheme surpasses traditional methods across all metrics, including minimum and mean ITR, lower error rates, and improved error statistics. Importantly, as illustrated by the virtual flashboard technique shown in Figures \ref{fig_seq} and \ref{fig_diag}, the underlying flashboard structure remains unchanged; it is the highlighting of the characters that varies with the particular scheme used. Hence, we believe that this should have a negligible impact on the increase in complexity and the subject's attention.

In Figures \ref{wscv_only} and \ref{ascv_only}, the improvement is proportional to the complexity of the language model, with GPT-2 standing out in performance. For instance, when Dijkstra’s algorithm is employed for word completion, the ITR gain increases to $63.6$ bits/minute, representing an increase of almost $15\%$. When GPT-2 is used in conjunction with Dijkstra's algorithm, the ITR further improves to $75.3$ bits/minute ($p < 0.001$), resulting in a net improvement of about $28\%$. This value is very close to the word performance limit (within $4\%$), suggesting that there is limited residual gain to exploit using large language models. However, one could explore predicting not only the next word, but also performing multi-word predictions to further increase ITR. In particular, the retry rate is generally much lower when large language models are utilized. Specifically, in Figures \ref{wscv_only}b and \ref{ascv_only}b, both the retry rate and the occurrence of outliers are considerably reduced, which is not surprising given the ITR gains. Moreover, regardless of the specific algorithm used, word completion algorithms provide substantial gains over schemes that rely solely on character prediction, consistent with previous work \cite{speier_spell}. It is worth noting that by applying smoothing, flashboard character probabilities for 17 biwords in the DOI dataset that were out of vocabulary (OOV) from the biword language model were obtained by transitioning to the word and trigram models.

Across subjects (ASCV), performance (Figure \ref{ascv_only} a) shows a similar trend to WSCV. However, compared to WSCV, more subjects exhibit lower ITRs and higher error rates (Figure \ref{ascv_only}b). The diagonal character flashboard showed a $7.6\%$ improvement over the regular flashboard with random flashing, while Dijkstra's and GPT-2 word completion provide $35.6\%$ and a massive $76\%$ gain, respectively. Notably, there is a significant improvement in the retry rate when word completion is employed. Interestingly, the gains from word prediction are not dependent on the language model, although the choice of the specific model does influence the relative gains.

Retrofitting, which refers to enhancing standard schemes with word prediction, increases the mean ITR to within $2\%$ of the best-in-class performance (diagonal+GPT-2) for both within-subject and across-subject analyses (Figure \ref{wscv_ascv}). This result is particularly insightful, as it suggests that traditional approaches can be preserved and upgraded with a word completion technique based on an advanced language model, alongside smoothing, to achieve near-optimal performance. It also highlights the substantial gains that can be achieved with powerful language models, demonstrating that word completion using the model offers significantly greater improvements compared to character-level flashboard enhancements.

The primary limitation of this study lies in the assumptions made during the simulation. Like all offline studies, it assumes that neural signals are not influenced by the presence of feedback. Although this is a strong assumption, it is supported by previous research showing consistent performance in online and offline analyses \cite{speier2}. Several factors render online testing impractical for this study. The model proposed aims to enhance system flexibility, accommodating rare and out-of-vocabulary words while testing system components that could previously only be demonstrated online. Furthermore, it reduces the scope of tedious BCI experimental evaluations by optimizing parameters over long simulation datasets. In our earlier online studies, subjects typically chose short sentences composed of common English words, which would not effectively showcase the value of the proposed models. While we could select a passage for our subjects, imposing a sentence with rare and/or OOV words would likely bias the analysis in favor of our model. Therefore, we determined that the best approach to evaluate our proposed model's performance was to select a long passage separate from our training corpus, increasing the likelihood of encountering rare or OOV words at a rate more consistent with general English text.

Since this document is too lengthy to realistically be typed during a BCI session, we designed the simulation environment to represent how the system could function for a large population of BCI subjects. While we acknowledge that simulation is an imperfect representation of BCI performance, we point out that in all our previous studies, offline performance gains from NLP and ML improvements have been sustained in online sessions \cite{speier2}, \cite{speier3}. Although online experiments are undoubtedly important, we firmly believe that this offline simulation is sufficient for the current study, with online analysis reserved for a future longitudinal study that can provide a larger body of testing output text.

While we consider GPT-2 a powerful language prediction tool, there are certainly contextual limitations regarding language generation \cite{openai}. In fact, language was not learned solely through word prediction. However, advancements in GPT-based language models, such as GPT-3 \cite{gpt3}, mitigate these limitations, further enhancing its value. Thus, we do not view this as a fundamental limitation of the results in this paper; rather, it opens avenues for experiments with new models that are more context-sensitive.

\section{Conclusion and future directions}
Language models that enhance the speed and accuracy of classification in the P300 speller can also be effectively used to create whole-word suggestions for predictive spelling. We introduced new techniques for flashboard construction based on probabilistic modeling. Combined with word-completion algorithms, typing speed is significantly improved compared to current random highlighting techniques. Using real-world data from $78$ subjects, simulations with a large dataset showed an increase in typing speed (over standard schemes) in within-subject training from $51.9$ bits/minute to $75.4$ bits/minute. An across-subject classifier methodology was also introduced, eliminating the need for subject-specific calibration. The same ITR metric improved from $33.9$ bits/minute to $59.7$ bits/minute with across-subject training and word prediction. Additionally, the error or retry rate decreased, closing the gap between theory and practice.

Further improvements may be achievable through dynamic thresholding techniques, allowing threshold settings to adapt based on subject responses, which will be the focus of future work.

\end{document}